\documentclass[superscriptaddress, twocolumn, 11pt, pre,floats]{revtex4-1}
\pdfoutput=1

\usepackage{amsmath}    
\usepackage{graphicx}   
\usepackage{verbatim}   
\usepackage{color}      
\usepackage{hyperref}   
\usepackage{amsmath,amssymb}
\begin{document}

\title{
Robust Regression for Automatic Fusion Plasma Analysis
based on Generative Modeling
}

\author{K. Fujii}
\affiliation{
Department of Mechanical Engineering and Science, \\
Graduate School of Engineering, Kyoto University, \\
Kyoto 615-8540, Japan,\\
Email: fujii@me.kyoto-u.ac.jp}
\author{C. Suzuki}
\affiliation{National Institute for Fusion Science, \\
Gifu 509-5292, Japan}
\author{M. Hasuo}
\affiliation{
Department of Mechanical Engineering and Science, \\
Graduate School of Engineering, Kyoto University, \\
Kyoto 615-8540, Japan,\\
Email: fujii@me.kyoto-u.ac.jp}

\date{\today}

\begin{abstract}
  The first step to realize
  automatic experimental data analysis for fusion plasma experiments
  is fitting noisy data of temperature and density spatial profiles, which are obtained routinely.
  However, it has been difficult to construct algorithms that fit
  all the data without over- and under-fitting.
  In this paper,
  we show that this difficulty originates from the lack of knowledge of the probability distribution that the measurement data follow.
  We demonstrate the use of a machine learning technique
  to estimate the data distribution and to construct an optimal generative model.
  We show that the fitting algorithm based on the generative modeling outperforms
  classical heuristic methods in terms of the stability as well as the accuracy.

  keywords: automatic analysis, Beyesian method, neural network, variational inference
\end{abstract}

\maketitle
\newcommand{\argmin}{\mathop{\mathrm{argmin}}\limits}
\newcommand{\argmax}{\mathop{\mathrm{argmax}}\limits}
\newcommand{\bvec}[1]{\mathbf #1}

\section{Introduction}
A variety of analysis methods for fusion plasma experiments have been developed,
such as
dynamic transport analysis
\cite{Takeiri2017,yokoyama2017,Highcock2007},
linear and nonlinear turbulence growth rate analysis
\cite{Ritz1988},
and others.
The automatic analysis of experimental data with these methods,
trial-by-trial, would help researchers to perform further analysis,
possibly resulting in acceleration of fusion research.
Because most of the analysis methods require spatial gradients of the
temperature and density profiles in the plasma,
the first step to realize automatic data analysis is
fitting the noisy measurement data with a smooth function.

\begin{figure*}
\centering
\includegraphics[width=6in]{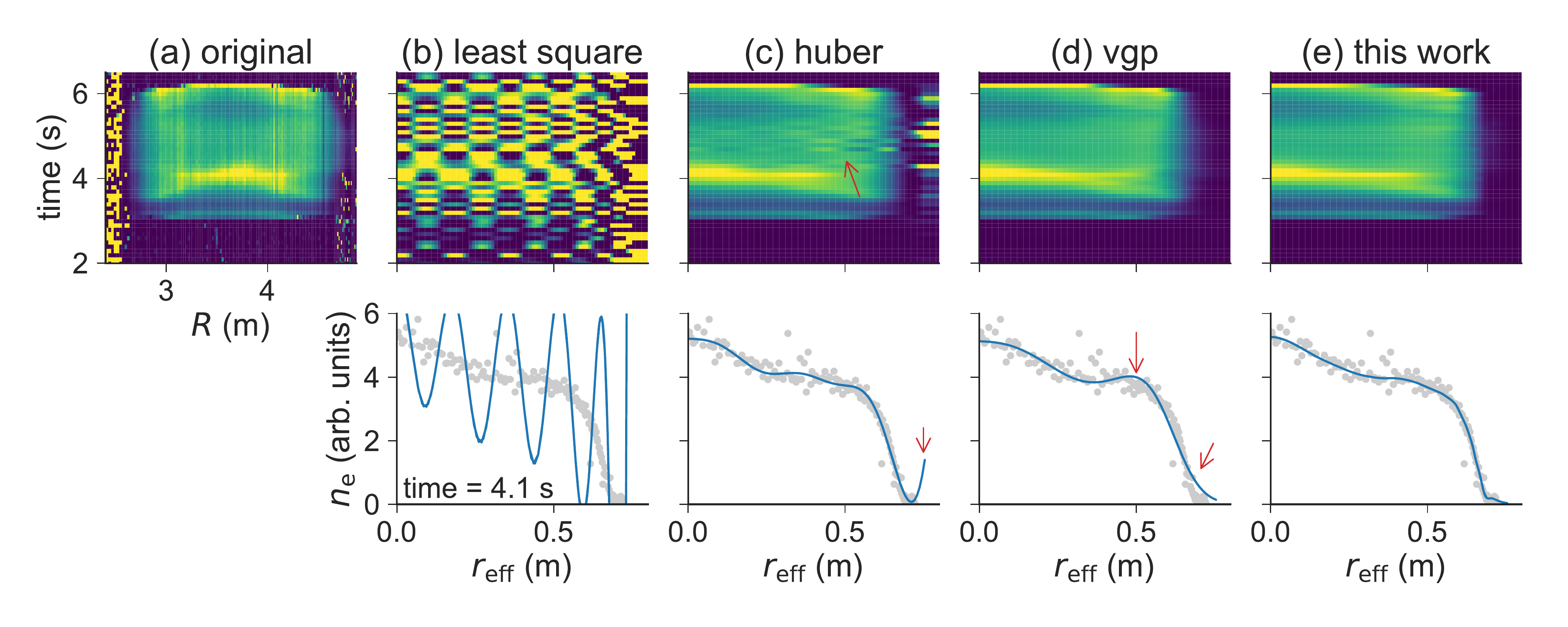}
\caption{
\label{fig:conventional_ne}
(a) Typical example of $N_\mathrm{e}$ data measured with Thomson scattering in LHD,
as a function of the major radius $R$ and the experimental time.
The scatter near the edge regions $R < 2.5$ m and 4.8 m $< R$ are  much larger
than the other region.
(b)--(e)
Fit results by several methods.
The top figures show the result
as a function of the minor radius $r_\mathrm{eff}$ and the experimental time.
The bottom figures show the data obtained at 4.1 s (markers) and
the corresponding fit result (solid curves).
(b)
Result with the least squares method with Gaussian basis.
This method is significantly affected by outliers,
resulting in an almost meaningless distribution.
(c)
Result with the Huber regression method with Gaussian basis.
Although the influence of the outliers is significantly reduced compared with those by
the least squares method, some artifacts can be seen in edge regions
(indicated by an arrow in the bottom figure).
Additionally, an oscillation structure is also seen in some time slices,
as shown by an arrow in the top figure.
(d)
Variational Gaussian process (VGP) regression with student-$t$ likelihood.
This method is found to be more robust than the Huber regression.
However, this method sometimes neglects the steep gradient at the edge regions
as indicated by an arrow in the bottom figure.
(e)
Fit results by this work.
Our method is as robust as the variational Gaussian process method but still
retains the detailed structure near the edge regions.
The details are discussed in Section \ref{subsec:conventional}.
}
\end{figure*}

\begin{figure*}
\centering
\includegraphics[width=6in]{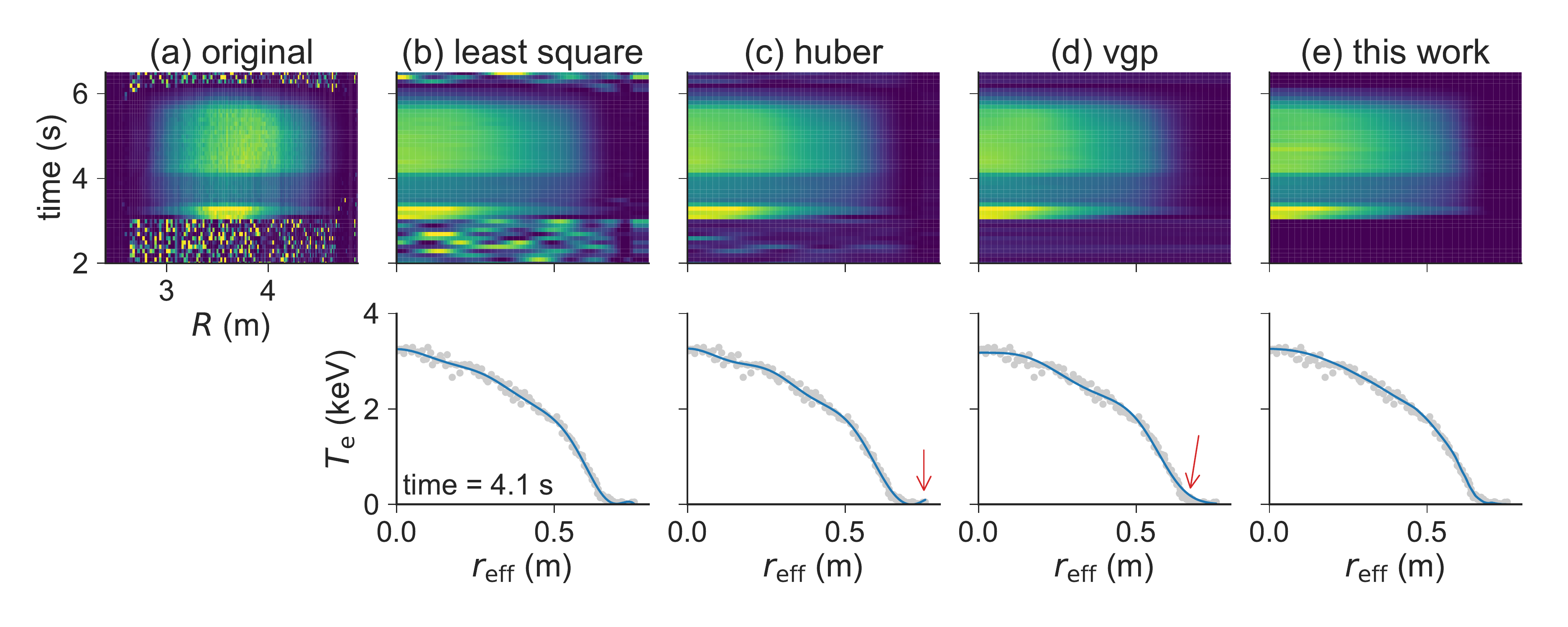}
\caption{
\label{fig:conventional_te}
Similar figure to Fig. \ref{fig:conventional_ne} but for $T_\mathrm{e}$ data.
There are fewer outliers in the edge regions than in the $N_\mathrm{e}$ data.
(b)--(d) Conventional methods including the least squares show the reasonable results, in 3 -- 6 s,
where the plasma discharge is taking place.
However, there are artifacts, even when there are no plasma (2--3 s and 6--7 s.),
in the conventional results.
Additionally, the steep gradient near the edge region ($r_\mathrm{eff} \approx 0.6$ m)
is averaged out in the conventional robust methods (Huber and VGP, as indicated by arrows).
}
\end{figure*}

Fitting measurement data is one of the first examples
of many basic textbooks in statistics,
but there are several differences in
\textit{real-world} data, particularly for fusion experiments.
As an example, in Figs. \ref{fig:conventional_ne} and \ref{fig:conventional_te},
we show typical spatial distributions of electron temperature ($T_\mathrm{e}$)
and density ($N_\mathrm{e}$) obtained by Thomson scattering in
the Large Helical Device (LHD) project \cite{Yamada2003, Yamada2010}.
The first difference from the textbook example is that
the noise distribution in the data is unknown.
As can be seen in Figs. \ref{fig:conventional_ne} and \ref{fig:conventional_te},
there are certain outlying points, suggesting non-Gaussian noise in the data.
Historically many heuristic methods have been proposed
to avoid the influence of outliers \cite{Cleveland2012, Andrews1974},
such as the median filter and the Huber regression \cite{Huber2009}.
These methods have been used in many fields of science,
but they are not always effective.
As shown in Figs. \ref{fig:conventional_ne} and \ref{fig:conventional_te},
these methods occasionally treat important features in the data as outliers
and vice versa.
Therefore, human supervision has been necessary for the profile fitting.

The second difference is that
the theoretical forms of these profiles (latent functions) are also unknown.
If we use a too complex model (e.g. a model with many parameters), the fit curve will be too much affected by the noise and often show oscillating structure, i.e., over-fitting.
Such an oscillation due to the over-fitting can be seen Fig. \ref{fig:conventional_ne}(b) as indicated by an arrow.
On the other hand, if we use too simple model (e.g. a model with only a few parameters), the fit curve does not represent the important structure that the data has, i.e., under-fitting, as can be seen in Fig. \ref{fig:conventional_ne}(c)

In order to reduce biases introduced by using inappropriate functions,
non-parametric fitting methods, such as Gaussian process regression,
have been proposed \cite{Rasmussen2006}
and have also been applied in fusion plasma science \cite{Chilenski2015}.
Although the profile data we want to fit is obtained routinely
and there is a certain similarity among them,
such a non parametric method is blinded by the similarity,
i.e., the prior distribution assumed in the method
is biased from the data distribution.
The details of this method will be discussed in Section \ref{subsec:conventional}.

In the next section,
we theoretically show that such under- and over-fittings are caused by
the discrepancy between the data distribution and a generative model
implicitly assumed in the analysis.
Next, we propose to learn the generative model from a vast amount of data,
with the aid of machine learning techniques.
Finally, we demonstrate an automatic regression analysis
with the optimized generative model.



\section{Conventional Robust Analyses}

The simplest method to fit experimental data
$\mathbf{y} = \{y_i | i=0, \hdots, N\}$ is
the least squares method,
\begin{equation}
  \label{eq:least_square}
  \argmin_\theta \left[\sum_i^N
    (y_i - f(\theta)_i)^2
  \right]
\end{equation}
where $\mathbf{f}(\theta) = \{f(\theta)_i | i = 0, \hdots, N\}$
is the latent function,
which is manually chosen before the analysis.
The function is usually parameterized by adjustable parameters $\theta$,
and the fitting is carried out by optimizing $\theta$.
The least squares method assumes Gaussian noise with homogeneous variance.
However, the noise distribution frequently deviates from homoscedastic Gaussian.
In particular, when the distribution has a heavy tail,
i.e., when there are outliers,
the least squares method performs poorly.
The plots in Figs. \ref{fig:conventional_ne} (b) and \ref{fig:conventional_te} (b)
show the results of
the least squares fitting with Gaussian basis functions.
The results are significantly affected by the outliers around the edge region $r_\mathrm{eff} \approx $ 0.7 m.

Many algorithms have been proposed to obtain more results against outliers \cite{Huber2009, Hardle, Hardle1984, Bioinformatics}.
One typical method is Huber regression,
which minimizes the following error function
\begin{equation}
  \label{eq:huber}
  \argmin_\theta \sum_i^N
    L_\delta(y_i - f(\theta)_i)
\end{equation}
with
\begin{equation}
  \label{eq:huber_loss}
  L_\delta(x)=
    \begin{cases}
      \frac {1}{2}{x^{2}}&{\text{for }}|x|\leq \delta\\
      \delta (|x|-{\frac {1}{2}}\delta )&{\text{otherwise}}
    \end{cases},
\end{equation}
where $\delta$ is a hyperparameter.
Points deviating by more than $\delta$ are less penalized than in the least squares method.
This is expected to be robust against outliers.

From the Bayesian statistical point of view,
these methods are equivalent to the point estimation of the following
posterior distribution
\begin{equation}
  \label{eq:bayes_ls}
  p(\theta | \mathbf{y}, \mathcal{M}) =
  p(\mathbf{y} | \mathbf{f}(\theta), \mathcal{M})
  p(\theta | \mathcal{M}),
\end{equation}
where $p(\theta | \mathcal{M})$ is the prior distribution
of $\theta$, which is assumed as a uniform distribution here.
$p(\mathbf{y} | \mathbf{f}(\theta), \mathcal{M})$ is the likelihood,
which is the conditional probability of obtaining the particular data
$\mathbf{y}$ with the given latent value $\mathbf{f}(\theta)$.
The equivalent likelihood for the least squares method is the homoscedastic Gaussian,
$p(\mathbf{y} | \mathbf{f}(\theta), \mathcal{M}^\mathrm{sq}) =
\prod_i^{N}\mathcal{N}(y_i | f(\theta)_i, \sigma)$,
where $\mathcal{N}(x | \mu, \sigma)$ is
the normal distribution of random variable $x$ with mean $\mu$ and variance $\sigma$.
Huber regression assumes
$p(\mathbf{y} | \mathbf{f}(\theta), \mathcal{M}^\mathrm{Huber}) =
\prod_i^{N}\frac{1}{Z}\exp\left[-L_\delta(y_i - f(\theta)_i)\right]$,
where $Z$ is the normalization factor.
As such, different robust regression methods exhibit the different likelihood
form.

Most of the methods assume homoscedastic noise,
i.e. the noise variance (or its statistical quantity) is assumed to be uniform
over all the data points.
However, many advanced measurement data exhibit
complex dependencies of the noise variance \cite{Fujii, Fujii2017}.
A typical example of this is Thomson scattering,
which measures the electron temperature and density from the spectral shape
of scattered light.
The measurement accuracy (i.e., the inverse of the noise amplitude) is lower
for lower-density plasma,
because of the lower scattered light intensity.
Additionally, if the temperature is much higher or lower than the system's optimum,
the accuracy also decreases.
The heteroscedasticity of the noise can be seen in
Fig. \ref{fig:conventional_ne}, where the noise in the edge region
(where the temperature and density are low) are larger than the other region.
Owing to the discrepancies between the assumed and real situations,
the fitting result becomes unsatisfactory.

\section{Theoretical Perspective of Robust Regression Analysis}

Let us consider measurement data $\mathbf{y}$ that are being obtained routinely.
These measurement data have been obtained previously and will be also obtained
in the future.
We consider fitting a particular ($k$-th) set of data $\mathbf{y}^{(k)}$ by a model
$\mathcal{M}$, which has a set of adjustable parameters $\theta$.
From the Bayesian statistics perspective,
the fitness of model $\mathcal{M}$ can be measured
by the logarithm of the marginal likelihood,
\begin{equation}
  \log p(\mathbf{y}^{(k)} | \mathcal{M}) =
  \log \int p(\mathbf{y}^{(k)} | \mathbf{f}(\theta), \mathcal{M})\;
  p(\theta | \mathcal{M}) \;
  d \theta,
\end{equation}
where $p(\theta | \mathcal{M})$ is the prior distribution of $\theta$.

One of the reasonable measures of the robustness of model $\mathcal{M}$
is this logarithmic marginal likelihood for the data
that will be obtained in the future,
$\mathbf{y}^{(k+1)}, \mathbf{y}^{(k+2)}, \hdots $.
Let $p(\mathbf{y})$ be the probability distribution that generates these data.
The most robust model should maximize
the expectation of the logarithmic marginal likelihood,
\begin{align}
  \label{eq:expectation_kl}
  \mathcal{M}^\mathrm{best} & =
    \argmax_\mathcal{M}\left[
    \mathop{\mathbb{E}}_{p(\mathbf{y})}
    \left[\log p(\mathbf{y} | \mathcal{M}) \right]
    \right] \\
  & =
  \argmax_\mathcal{M}\left[
  \int p(\mathbf{y}) \;
  \log p(\mathbf{y} | \mathcal{M})\; d\mathbf{y}
  \right].
\end{align}
Because the term in the square brackets is identical to the negative of
the Kullback-Leibler (KL) divergence
between $p(\mathbf{y})$ and $p(\mathbf{y} | \mathcal{M})$, except for a constant,
this equation can be summarized as
\begin{align}
  \label{eq:data_kl}
  \mathcal{M}^\mathrm{best} & =
  \argmin_\mathcal{M} \mathrm{KL}\bigl[
    p(\mathbf{y}) \bigm\vert\bigm\vert p(\mathbf{y} | \mathcal{M}) \bigr],
\end{align}
where
$\mathrm{KL}[p(x) \vert\vert q(x)] = \int{p(x) \log \frac{p(x)}{q(x)} dx}$
is the KL divergence,
which is a measure of the distance between two distributions, $p(x)$ and $q(x)$.
$p(\mathbf{y} | \mathcal{M})$ can be viewed as a \textit{generative model},
which is a modeled probability distribution that generates data $\mathbf{y}$.

The first conclusion we can draw from Eq. \ref{eq:data_kl} is that
\textit{the robustness of the model depends on the data distribution}
$p(\mathbf{y})$.
Therefore,
there are no superior models that are always best for any type of data
distribution.
This might be a reason why there have been so many robust algorithms reported.
The second conclusion is that
\textit{the robustness is determined not only by the likelihood but
also by the function form and the prior distribution.}
This is because our generative model $p(\mathbf{y} | \mathcal{M})$ is derived
from the product of the likelihood and the prior.

\section{Proposed Method \label{sec:model_opt}}
From the above discussion, the best model depends on the data distribution.
In this work, we propose constructing an optimum generative model,
i.e., the likelihood, prior, and the function form,
from a vast amount of the previous data.
The overall strategy is summarized below.

\begin{enumerate}
  \item \label{item:prepare}
  Prepare the dataset and divide it into training, validation, and testing data sets.
  $p(\mathbf{y})$ is approximated by them.

  \item \label{item:parameterize}
  Parameterize the likelihood, prior, and latent function with neural networks,
  with weights $\Theta^\mathrm{lik}$, $\Theta^\mathrm{pri}$ and
  $\Theta^\mathrm{fun}$, respectively.
  Note that the form of the parameterization,
  such as the network structure and function forms, are treated as
  hyperparameters $\Omega$.

  \item \label{item:training}
  Estimate the best model $\mathcal{M}^\mathrm{best}$ by
  optimizing $\Theta^\mathrm{lik}$, $\Theta^\mathrm{pri}$, and
  $\Theta^\mathrm{fun}$ with respect to Eq. \ref{eq:data_kl} for the training set.
  Calculate the values of Eq. \ref{eq:data_kl} for the validation data.
  Compare these values for several hyperparameter settings $\Omega$
  and choose one of them.
\end{enumerate}

\subsection{Data preparation}

We apply our proposed strategy to
$T_\mathrm{e}$ and $N_\mathrm{e}$ data provided by the
Thomson scattering system in LHD.
In the Thomson scattering system,
the $T_\mathrm{e}$ and $N_\mathrm{e}$ values are
estimated from the scattered light spectral profiles.
Because this measurement is based on non linear signal processing,
it is difficult to evaluate the uncertainty of the estimated values.
The system only provides an estimate of the uncertainty based on a covariance matrix.
Typical results from Thomson scattering are shown in Figs.
\ref{fig:conventional_ne} and \ref{fig:conventional_te}.

Although the original measurement position is the plasma major radius ($R$),
we used the result of the magnetic coordinate mapping \cite{Suzuki2013},
i.e., the measurement position is already mapped to the plasma minor radius
($r_\mathrm{eff}$).

We need to approximate $p(\mathbf{y})$ from the existing finite data samples
to optimize $\mathcal{M}$ with respect to Eq. \ref{eq:data_kl}.
If there were no (long-term) temporal evolutions of the data distribution,
the existing data would be regarded as random samples from $p(\mathbf{y})$.
However, experimental conditions, such as the instrumental calibration,
and noise characteristics, 
may vary in time.
In order to robustly estimate $p(\mathbf{y})$,
we prepared data obtained in experiment campaigns during the period 2011 -- 2017
(six campaigns)
and divided these data samples into
training data (obtained in 2013--2017),
validation data (obtained in 2012), and
testing data (obtained in 2011).
We trained the model,
i.e., we optimized $\Theta^\mathrm{lik}$, $\Theta^\mathrm{pri}$, and
$\Theta^\mathrm{fun}$,
against the training data.
The hyperparameter tuning is carried out with the validation data.
The results will be shown with the testing data.



\subsection{Parameterization}
\begin{figure}[h!]
\centering
\includegraphics[scale=1.0]{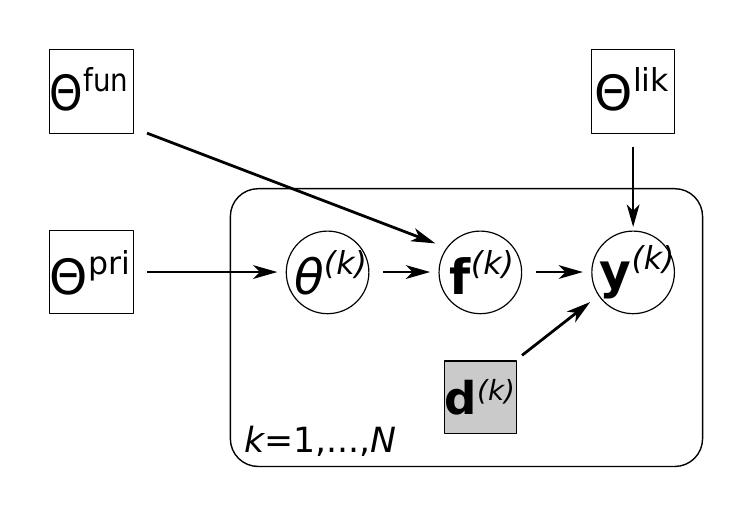}
\caption{
\label{fig:graphical_model}
Graphical representation of our generative model.
$\Theta^\mathrm{pri}$ parameterizes the prior distribution
$p(\theta | \Theta^\mathrm{pri})$ and $\theta$ for experiment $k$
is generated from this distribution.
The latent function values $\mathbf{f}^{(k)}$ for this experiment are
determined from Eq. \ref{eq:function_form}, where the basis function $\phi$
is parameterized by $\Theta^\mathrm{fun}$.
Data $\mathbf{y}^{(k)}$ are generated from the latent function value
$\mathbf{f^{(k)}}$ and other data $\mathbf{d}^{(k)}$
according to the likelihood distribution
parameterized by $\Theta^\mathrm{lik}$.
}
\end{figure}

In this section,
we present the function form, prior distribution, and likelihood form
adopted for this work.
A graphical representation of our overall model is summarized in Fig. \ref{fig:graphical_model}.

\subsubsection{Function form}
The function form to be used for the fitting should have two groups of parameters,
i.e., the local parameter $\theta$ and the global parameter $\Theta^\mathrm{fun}$.
The local parameter represents the individual characteristics of the data,
whereas the global parameter expresses the similarity among them.
For our particular purpose, we use the following generalized linear model
mapped by an exponential function to the positive space,
\begin{equation}
  \label{eq:function_form}
  f(r, \theta) = \exp\left[\sum_i^M \theta_i\;
                 \phi_i(r | \Theta^\mathrm{fun}) + \theta_\mathrm{bias}\right].
\end{equation}
The function is represented by a linear combination of $M$ basis functions
with the coefficient $\theta$ (local parameters).
$\phi_i(r | \Theta^\mathrm{fun})$ is constructed by neural networks,
the weights of which are $\Theta^\mathrm{fun}$ (global parameters).
Note that the basis function shapes are common for all the data whereas
$\theta$ is specific to each dataset.

\subsubsection{Prior distribution}
For the prior distribution of $\theta$,
we adopt independent normal distributions,
\begin{equation}
  \label{eq:prior}
  p(\theta_i | \Theta^\mathrm{pri}) = \mathcal{N}(\theta_i | \mu_i, 1),
\end{equation}
the mean $\mu_i$ is common for all the data,
and therefore $\mu_i$ is regarded as a global parameter $\Theta^\mathrm{pri}$.
Note that this model has some similarities with Gaussian process regression,
which has been used in the profile fitting with Bayesian methods \cite{Rasmussen2006,Chilenski2015}.

\subsubsection{Likelihood form}
In order to model the heteroscedasticity of the noise and its deviation from
Gaussian,
we adopt the following parameterization,
\begin{equation}
  \label{eq:likelihood}
  p(y_i | f, \Theta^\mathrm{lik}) =
  \mathcal{S}t\bigl(y_i |
    f, \sigma(f, \mathbf{d}^{(k)} | \Theta^\mathrm{lik}),
    \nu(f, \mathbf{d}^{(k)} | \Theta^\mathrm{lik})\bigr),
\end{equation}
where $\mathcal{S}t(x | \mu, \sigma, \nu)$ is the student's $t$ distribution
for random variable $x$ with mean $\mu$, scale parameter $\sigma$, and
degree of freedom $\nu$.
For modeling the heteroscedasticity of the noise,
we assume that both the scale parameter and the degree of freedom depends on
the true function value $f$.
Additionally, there might be other effects that make the data more noisy.
For example, the noise amplitude of the sensor preamplifier
might have a dependency on the environmental temperature or humidity.
We assume that such information can be estimated from other experimental data
$\mathbf{d}^{(k)}$,
such as the estimated uncertainties of $T_\mathrm{e}$ and $N_\mathrm{e}$ provided
by the Thomson scattering team.
We parameterize these dependences by neural networks,
which map the input ($\mathbf{f}^{(k)}$ and $\mathbf{d}^{(k)}$)
to the scale parameter and the degree of freedom.
Because these relationships should be common for all the data,
their weights $\Theta^\mathrm{lik}$ are regarded as global parameters.

\subsection{Training}
We optimized the global parameters
$\Theta^\mathrm{lik}$, $\Theta^\mathrm{pri}$ and $\Theta^\mathrm{fun}$
against the training dataset,
\begin{widetext}
\begin{equation}
\begin{aligned}
  \label{eq:training}
  \Theta^\mathrm{lik}, \Theta^\mathrm{pri}, \Theta^\mathrm{fun}
  &= \argmin_{\Theta^\mathrm{lik}, \Theta^\mathrm{pri}, \Theta^\mathrm{fun}}
  \left[
  \frac{1}{N^\mathrm{train}}\sum_{\mathbf{y}^{(k)}\in \mathrm{Y^{train}}}
      -\log p(\mathbf{y}^{(k)} |
             \Theta^\mathrm{lik}, \Theta^\mathrm{pri}, \Theta^\mathrm{fun})
  \right] \\
  & = \argmin_{\Theta^\mathrm{lik}, \Theta^\mathrm{pri}, \Theta^\mathrm{fun}}
  \left[
    \frac{1}{N^\mathrm{train}}\sum_{\mathbf{y}^{(k)}\in \mathrm{Y^{train}}}
      -\log \int
      p(\mathbf{y}^{(k)} |
        \mathbf{f}(\theta | \Theta^\mathrm{fun}), \Theta^\mathrm{lik})\;
      p(\theta | \Theta^\mathrm{pri}) \;
      d \theta
  \right],
\end{aligned}
\end{equation}
\end{widetext}
where the expectation in Eq. \ref{eq:expectation_kl} is approximated
by a Monte-Carlo method.
We have made several approximations to carry out the optimization of the equation,
because Eq. \ref{eq:training} is not analytically tractable.
The details of these are included in Appendix \ref{app:approximation}.
Some hyperparameters, such as network structures, are optimized based on the
validation dataset.

\section{Results}
\begin{figure}
\begin{center}
\includegraphics[width=7cm]{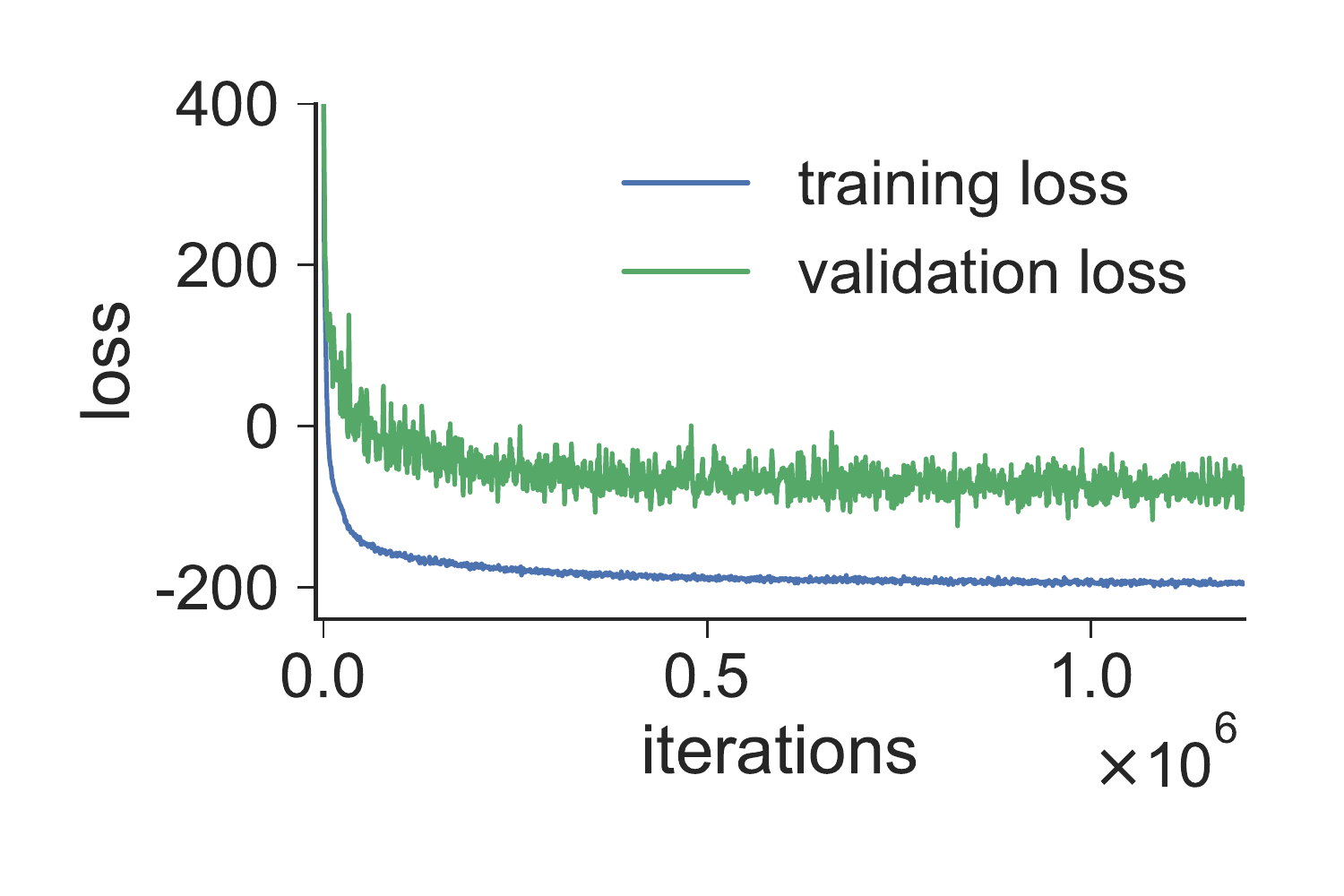}
\end{center}
\caption{
\label{fig:loss}
Temporal evolution of the loss value for the training data
 (blue curve, Eq. \ref{eq:training})
and that for the test data (green curve)
as a function of the iteration number.
}
\end{figure}

Figure \ref{fig:loss} shows the temporal evolutions of the approximated values of Eq. \ref{eq:training}.
These values for the training and test data sets converged in less than $10^6$ iterations.
There are some differences between the training and validation losses.
As we divided all the data into training, validation, and testing data sets by
time,
this difference indicates a temporal drift in the data distribution.
%
%
Therefore, it is important to fit not only the training data but also
the validation data.

\subsection{Typical results}

\begin{figure*}
\begin{center}
\includegraphics[width=12cm]{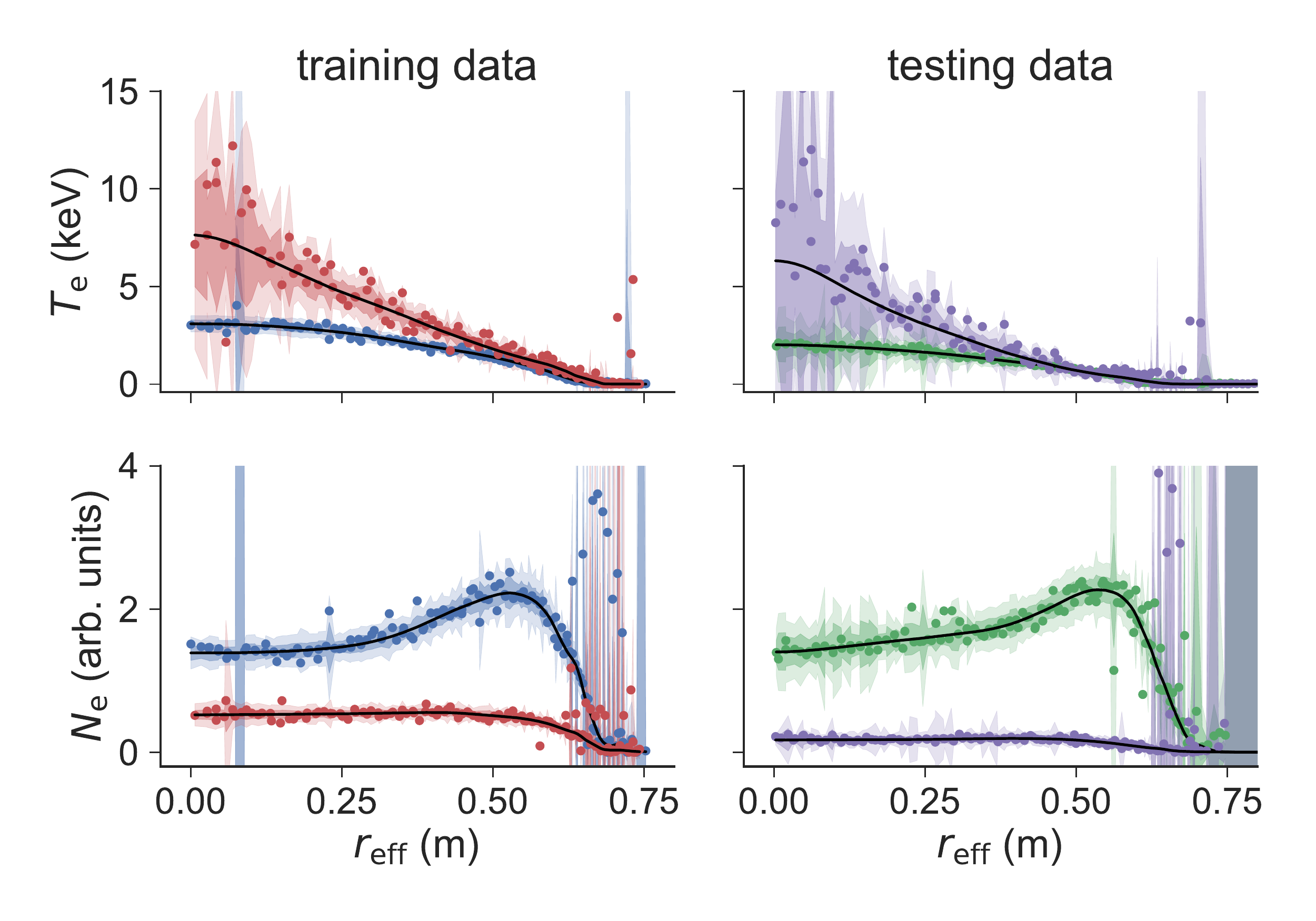}
\end{center}
\caption{
\label{fig:fit}
Typical data of $T_\mathrm{e}$ and $N_\mathrm{e}$ data measured by
LHD Thomson scattering system.
The figures on the left show training data and those on the right show
validation data.
Our fitting results by Eq. \ref{eq:function_form} are indicated by solid curves.
The shaded regions indicate the coverage probabilities of the likelihood
(Eq. \ref{eq:likelihood}),
where light and dark regions indicate 95\% and 65\% coverage probability, respectively.
Note that for a normal distribution,
the region with 95\% coverage probability is roughly twice as wide as the 65\% region.
}
\end{figure*}

Figure \ref{fig:fit} shows typical fitting results for
training and testing data.
The black curves show the medians of the posterior distribution of the latent functions,
whereas the dark and shaded areas show
the 65\% and 95\% regions of the likelihood.
Note that with a Gaussian distribution,
the 95\% region is nearly twice as wide as the 65\% region,
whereas with the student's $t$ distribution with a lower degree of freedom,
the difference increases.
Because we use not only the latent function value but also uncertainty data
provided by the diagnostic team to
predict the likelihood,
the shaded region is not smooth against the radial coordinate.

Most of the training data are within the 95\% region.
It can be seen that the region containing the outlying points is much
larger than for the ordinary points.
The distribution of the testing data looks slightly different from
the prediction,
i.e., certain points deviate further than the shaded region.
This indicates that the distributions of the training and validation datasets are slightly different.

\begin{figure}[h]
\begin{center}
\includegraphics[width=8cm]{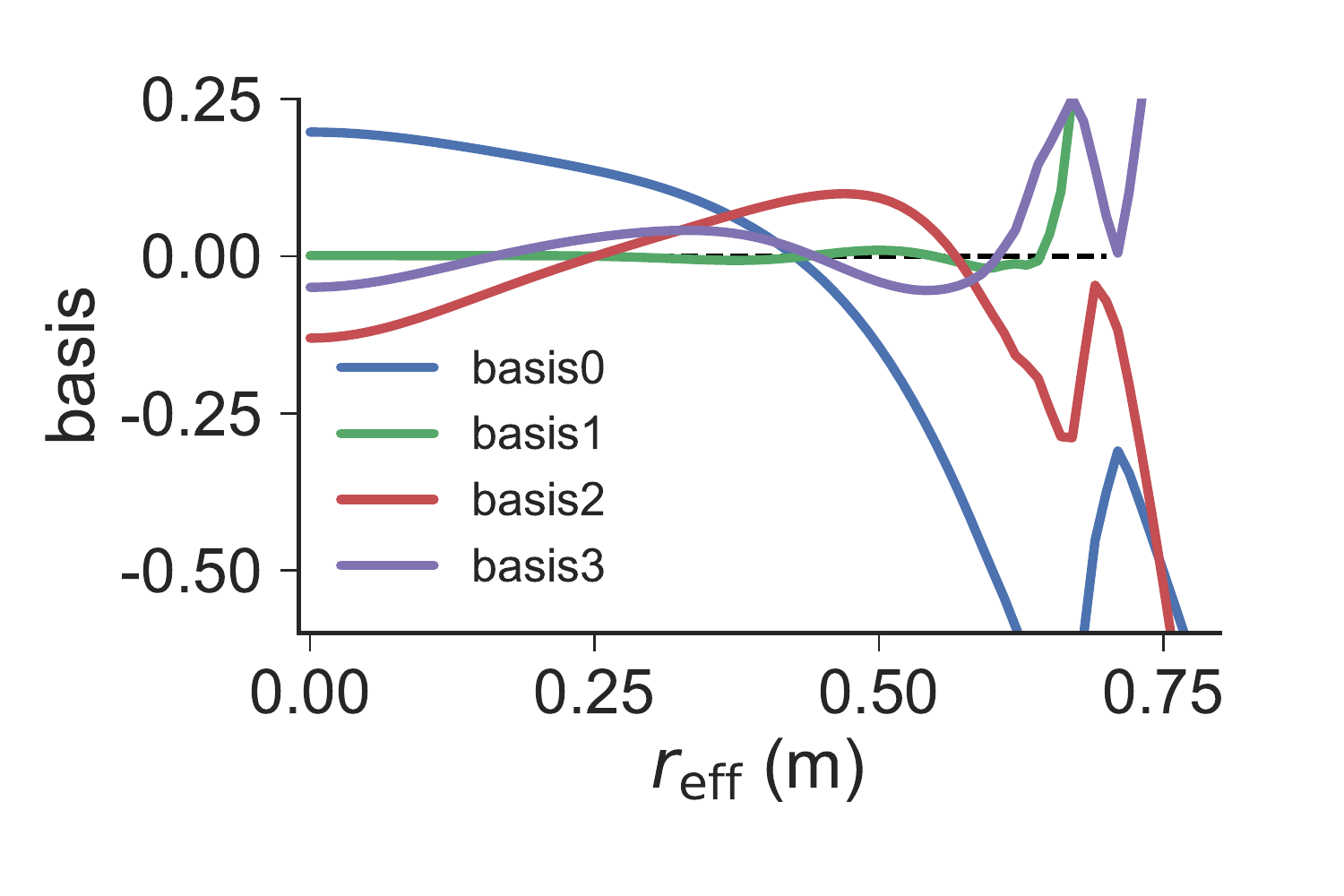}
\end{center}
\caption{
\label{fig:basis}
First four basis functions $\phi(r)$ for $T_\mathrm{e}$ and $N_\mathrm{e}$ profiles.
}
\end{figure}

Figure \ref{fig:basis} shows four of the basis functions
used both for the $T_\mathrm{e}$ and $N_\mathrm{e}$ profiles,
which are obtained from the optimization process.
Three basis functions ($\phi_0$, $\phi_2$, $\phi_3$) have different
spatial resolutions at $r < 0.6$,
where $\phi_0$ is the most smooth, and $\phi_3$ has an oscillation structure.
Sharp structures around $r \approx 0.7$ m can be seen in these basis functions.
They might represent the intrinsic magnetic island structure in LHD.
$\phi_1$ has a steep gradient at $r > 0.7$ m,
which is used to represent the steep gradient near the last closed flux surface.

All the basis functions have a low spatial resolution in the core region
($r_\mathrm{eff} < 0.6$ m) and a high resolution in the edge region ($r_\mathrm{eff} \approx 0.65$ m).
This characteristic is similar to the work by Chilenksi et al
\cite{Chilenski2015},
in which they heuristically adopted a Gaussian process kernel
with a smaller lengthscale around the edge region.

\begin{figure*}
\begin{center}
\includegraphics[width=16cm]{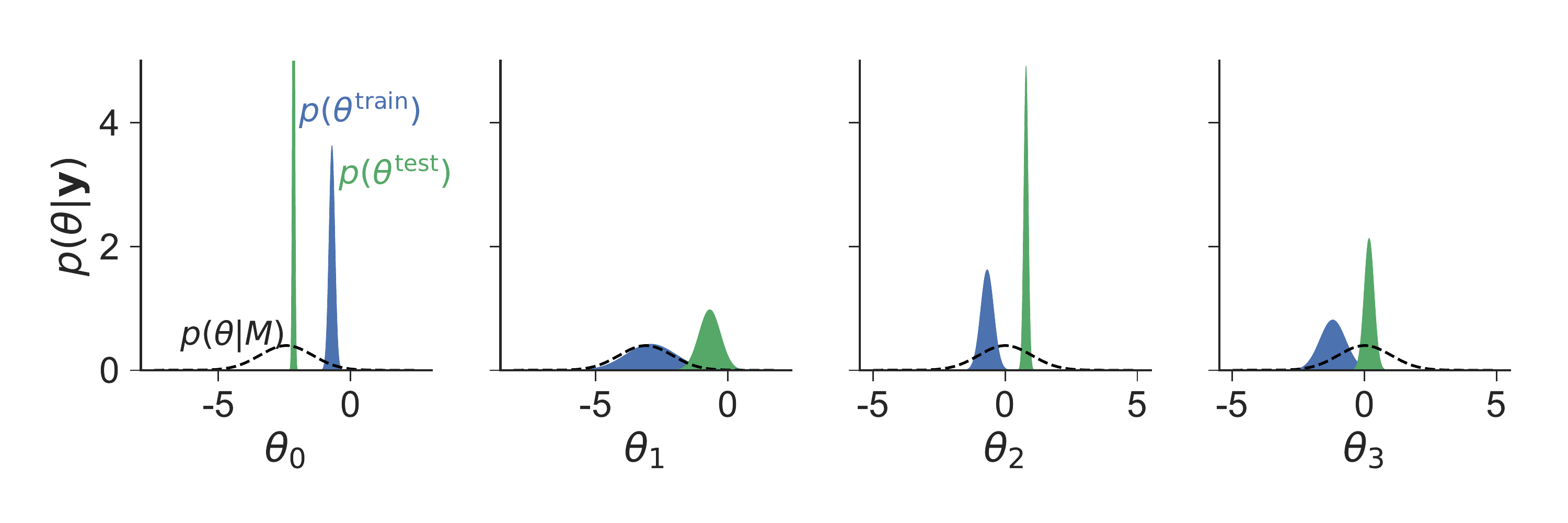}
\end{center}
\caption{
\label{fig:posterior}
Posterior distributions of first four latent parameters $\theta$.
The blue and green distributions correspond to
the posteriors of the training and testing data
(shown by the blue and green dots in Fig.\ref{fig:fit}), respectively.
The dotted curves show the prior distributions.
}
\end{figure*}

Figure \ref{fig:posterior}
shows the posterior distribution of the latent variable
$p(\theta | \mathbf{y}, \mathcal{M})$ corresponding to the data shown in
Fig. \ref{fig:fit}.
All the posterior distributions are narrower than the prior distribution
$p(\theta | \mathcal{M})$ (shown by dotted curves in Fig. \ref{fig:posterior}),
as we gained information from data $\mathbf{y}$.
The posterior distributions for $\theta_1$ are the broadest,
indicating that the basis function $\phi_1$ is less used for this data.

\begin{figure*}
\begin{center}
\includegraphics[width=12cm]{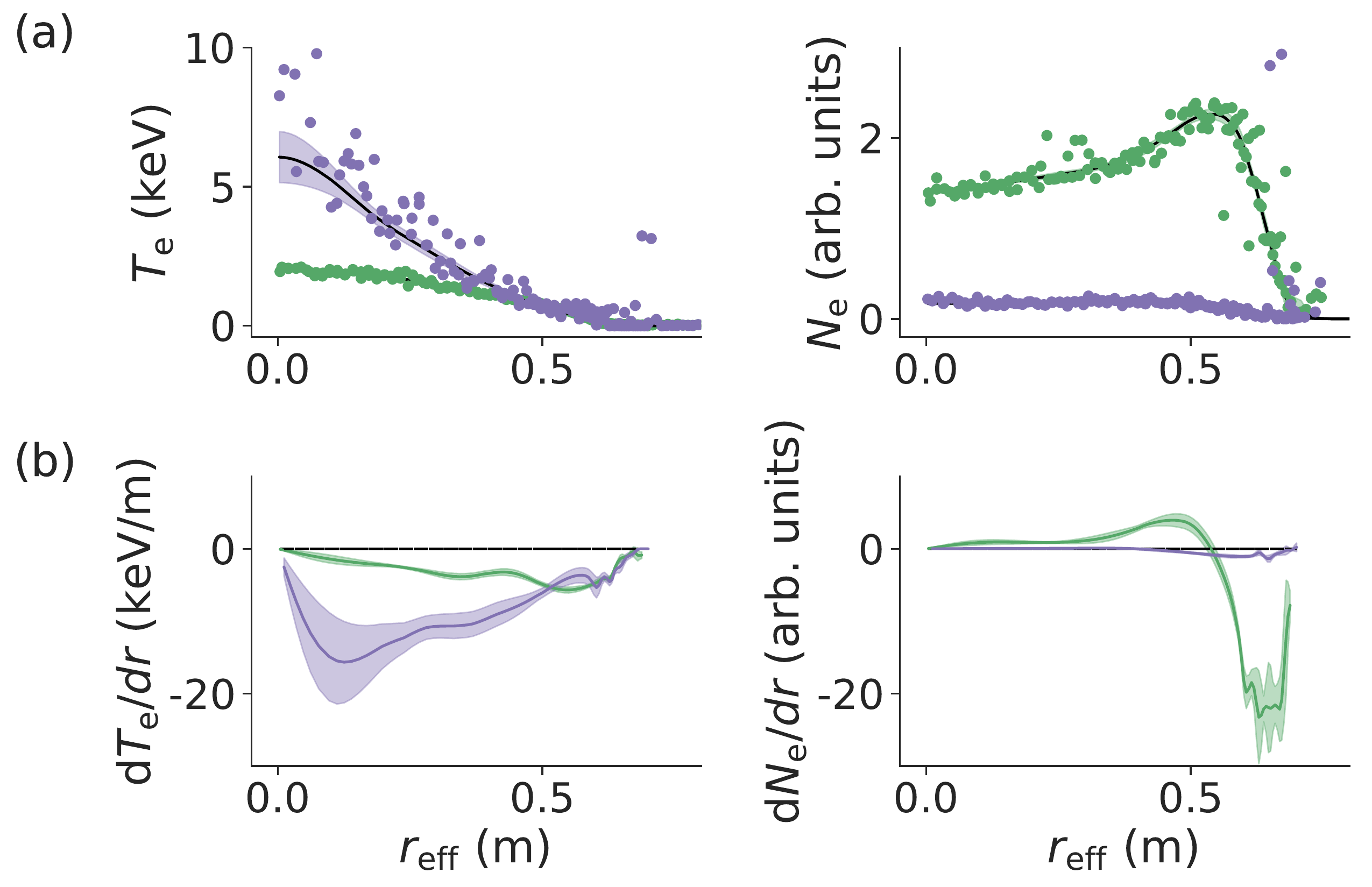}
\end{center}
\caption{
\label{fig:gradient}
(a) Posterior distributions of the fit curves for the testing data that are shown in the right pane in Fig. \ref{fig:fit}.
The mean is shown by the solid lines (which are the same those in Fig. \ref{fig:fit}) while $2 \sigma$ region of its uncertainty is shown by colored bands.
(b) The posterior distribution of the spatial gradients. The uncertainty ($2 \sigma$ region) is also shown by the colored band.
}
\end{figure*}

From $p(\theta | \mathbf{y}, \mathcal{M})$, the posterior distribution of the curve, $p(\mathbf{f} | \theta, \mathbf{y}, \mathcal{M})$, is also derived, which can be thought as the uncertainty of the estimated quantity.
In Fig. \ref{fig:gradient}(a), we show the posterior distribution for the curves corresponding to the testing data which is shown in Fig. \ref{fig:fit}.
Two $\sigma$ region of its uncertainty is shown by colored bands.
In Fig. \ref{fig:gradient}(b), we also show the spatial gradients for these data with their uncertainty.
Steep gradients around the edge region are quantified by the fit.

\subsection{Closeness of the learned and data distributions}

In this subsection, we visualize the closeness of our generative model
$p(\mathbf{y} | \mathcal{M})$ to the data distribution $p(\mathbf{y})$.
As described in Appendix \ref{app:closeness}, the following equality is satisfied
for Eq. \ref{eq:data_kl},
\begin{widetext}
\begin{equation}
  \label{eq:kl_ineq}
  \mathrm{KL}\bigl[
    p(\mathbf{y}) \bigm\vert\bigm\vert p(\mathbf{y} | \mathcal{M}) \bigr]
    =
    \int p(\theta) \mathrm{KL}\bigl[p(\mathbf{y}|\theta) \vert\vert
                                    p(\mathbf{y}|\theta, \mathcal{M})\bigr] d\theta
    +
    \mathrm{KL}\bigl[p(\theta) || p(\theta| \mathcal{M})\bigr].
\end{equation}
\end{widetext}
where $p(\mathbf{y} | \theta)$ and $p(\mathbf{y} | \theta, \mathcal{M})$
are defined as
\begin{equation}
  p(\mathbf{y} | \theta) = p(\theta | \mathbf{y}, \mathcal{M})
                           p(\mathbf{y}) / p(\theta)
\end{equation}
and
\begin{equation}
  p(\mathbf{y} | \theta, \mathcal{M}) = p(\theta | \mathbf{y}, \mathcal{M})
  p(\mathbf{y} | \mathcal{M}) / p(\theta | \mathcal{M})
\end{equation}
where $p(\theta | \mathbf{y}, \mathcal{M})$ is the posterior distribution
as shown in Fig. \ref{fig:posterior}.
$p(\theta | \mathcal{M})$ is the prior distribution of our generative model
\begin{equation}
  \label{eq:prior}
  p(\theta | \mathcal{M}) =
  \int p(\theta | \mathbf{y}, \mathcal{M})
        p(\mathbf{y} | \mathcal{M}) d\mathbf{y}
\end{equation}

\begin{figure*}
\begin{center}
\includegraphics[width=16cm]{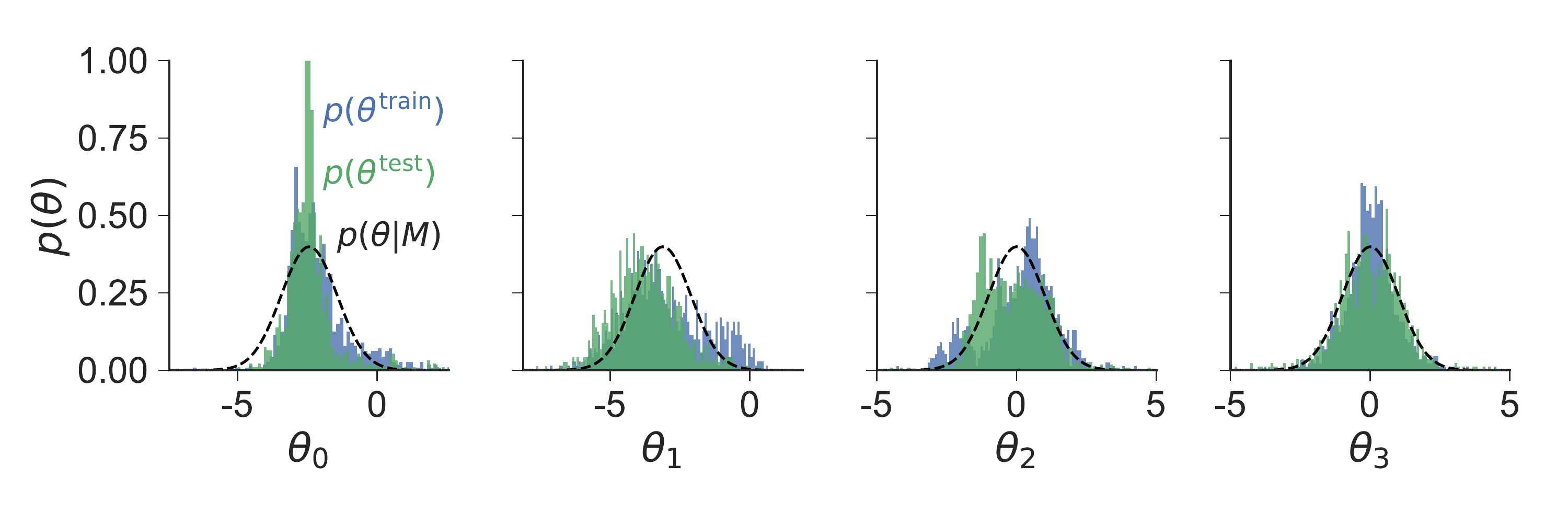}
\end{center}
\caption{
\label{fig:p_theta}
Prior distribution of the first four latent parameters $\theta$
$p(\theta | \mathcal{M})$ (dotted curves)
and the distributions of $\theta$ over the dataset $p(\theta)$
for the training (blue histograms) and validation (green histograms) datasets.
}
\end{figure*}

This relation states that the minimization of
$\mathrm{KL}\bigl[
  p(\mathbf{y}) \bigm\vert\bigm\vert p(\mathbf{y} | \mathcal{M}) \bigr]$
(Eq. \ref{eq:data_kl}),
requires the minimizations of
$\mathrm{KL}\bigl[p(\theta) \bigm\vert\bigm\vert p(\theta | \mathcal{M}) \bigr]$
and
$\mathrm{KL}\bigl[p(\mathbf{y} | \theta) \bigm\vert\bigm\vert
                   p(\mathbf{y} | \theta, \mathcal{M}) \bigr]$.
Figure \ref{fig:p_theta} shows $p(\theta)$ and $p(\theta | \mathcal{M})$
for the first four latent variables ($\theta_0, \hdots, \theta_3$).
In this figure,
we show $p(\theta)$ for both the training and testing datasets.
Although the latent variable distributions in the training and validation
datasets differ slightly,
both are close to our modeled prior distribution (dotted curves).

\begin{figure}[h]
\begin{center}
\includegraphics[width=8cm]{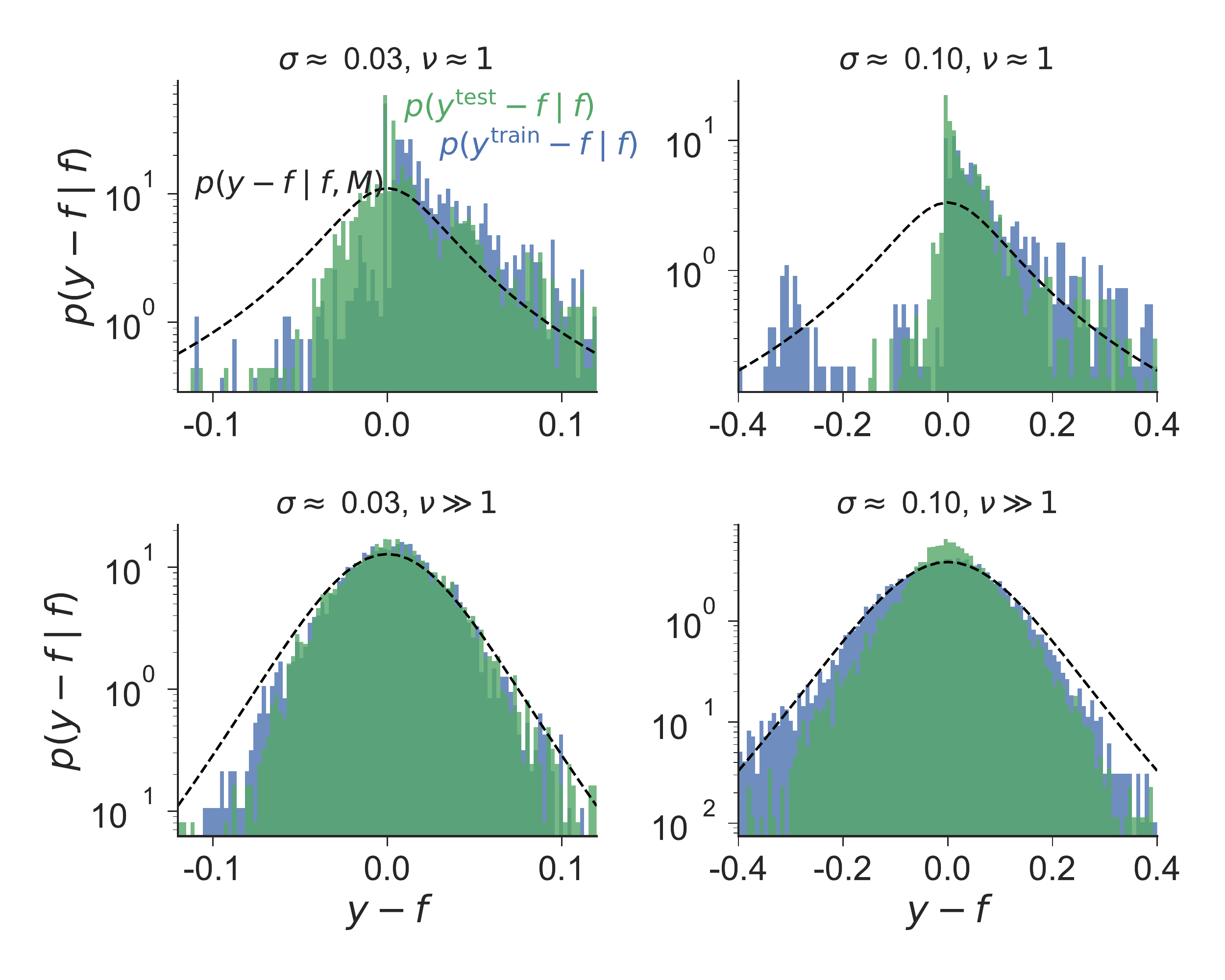}
\end{center}
\caption{
\label{fig:noise_pdf}
The estimated likelihood
$p(\mathrm{y} - \mathrm{f}(\theta) | \theta, \mathcal{M})$
(dotted curves) and
noise distribution $p(\mathrm{y} - \mathrm{f}(\theta) | \theta)$
over training (blue histograms) and
the validation (green histograms) datasets.
In order to clarify the heteroscedasticity of the data,
we show the four histograms,
conditioned by $\sigma(f(\theta)) \approx 0.03$ and $0.10$
(left and right figures, respectively)
and by $\nu \approx 1$ (close to Cauchy distribution, top figures) and
$\nu \gg 1$ (close to Gaussian distribution, bottom figures).
}
\end{figure}

Figure \ref{fig:noise_pdf} visualizes
the noise distribution (i.e., the likelihood $p(\mathbf{y} | \theta, \mathcal{M})$) of our generative model,
as well as the noise distribution in the training and testing datasets.
Because we model the likelihood by the student's $t$ distribution
(Eq. \ref{eq:likelihood}),
parameterized by $\sigma$ and $\nu$,
in this figure, we showed some typical cases of $\sigma$ and $\nu$.
Our likelihood model with $\nu \gg 1$ (lower panel)
is close to the true noise distribution.
The noise distribution with $\nu \approx 1$ (upper panel)
has a larger tail than the Gaussian distribution.
This is also consistent with our likelihood modeling
(the $t$ distribution with $\nu \approx 1$ has a much larger tail than the Gaussian distribution).
This indicates that our likelihood is less biased
than the simple Gaussian likelihood.

However, in the case with $\nu \approx 1$,
the noise distributes mainly on the positive side
whereas our assumption is symmetric.
This suggests that our generative model would be further improved by adopting
an asymetric likelihood.

\subsection{Comparison with conventional methods}\label{subsec:conventional}

In this subsection, we compare our results with those by conventional methods.
The data shown in
Figs. \ref{fig:conventional_ne}-\ref{fig:conventional_te} (a),
which are testing data
obtained for a typical LHD experiment (shotnumber \#111394),
are used as an example for the demonstration.
These figures show the temporal evolutions of the observed data.

The comparisons were made with
the least squares method with Gaussian basis
(Figs. \ref{fig:conventional_ne}-\ref{fig:conventional_te} (b)),
Huber regression with Gaussian basis
(Figs. \ref{fig:conventional_ne}-\ref{fig:conventional_te} (c)),
and
Gaussian process with student $t$ likelihood
(Figs. \ref{fig:conventional_ne}-\ref{fig:conventional_te} (d)).
For all the methods, we fit the data frame-by-frame,
i.e., each frame in the experimental data is treated independently.

Predictions with our method are shown in
Figs. \ref{fig:conventional_ne}--\ref{fig:conventional_te} (e).
It can be seen that our method is more robust and accurate
than the conventional fitting methods.
The details of the other methods are described in the rest of this subsection.

\subsubsection{Least squares fit}
We prepared 11 gaussian bases
with widths of 0.15 m and evenly located with 0.1 m intervals.
The fitting is performed according to Eq. \ref{eq:least_square}.
Because this method assumes homogeneous Gaussian noise,
which is extremely different from the actual noise,
the results are significantly affected by outliers
and they thus provide no meaningful information.

\subsubsection{Huber regression}
We used the same basis as above but minimize Eq. \ref{eq:huber}.
Its likelihood distribution (Eq. \ref{eq:huber_loss}) is closer to the actual
noise distribution than the homogeneous Gaussian,
and, therefore, the result passes through the center of the scattered data.
However, as shown by the arrows in
Figs. \ref{fig:conventional_ne}--\ref{fig:conventional_te} (c),
it sometimes predicts an oscillating structure that is not seen in the original data.

\subsubsection{Variational Gaussian process regression}
Gaussian process regression is one of the popular non parametric regression
methods, in which no function form is specified.
We omit the details of this method, as there are many detailed description
in the literature, e.g., \cite{Rasmussen2006, Opper2009}.
Although Gaussian noise distribution is typically assumed
for Gaussian process regression,
variational methods have been proposed to use different likelihoods
\cite{Opper2009}.

In this comparison, we adopt the student's $t$ distribution for the likelihood.
The degree of freedom is assumed as 3 \cite{Opper2009} and the noise scale
is estimated by the marginal likelihood maximization.
Additionally, we assume the Gaussian process in the logarithmic space.
In other word, we remapped the latent function by an exponential transform
such that the prediction always remains in the positive space,
as with our method.

The results are shown in
Figs. \ref{fig:conventional_ne}--\ref{fig:conventional_te} (d).
This method is more robust than Huber regression,
partly because it can estimate the function form from the data
and the student's $t$ distribution might be closer to the true noise distribution.
The detailed temporal evolution of the data is apparent.

This method does not consider the similarity among the multiple data.
For example,
typical $N_\mathrm{e}$ data has a steep gradient around the edge region.
However, this method does not distinguish between the steep gradient and outliers,
and it therefore predicts a gentler gradient in the edge retion
(as indicated by red arrows in the figures).

\subsection{More results}

\begin{figure*}
\begin{center}
\includegraphics[width=12cm]{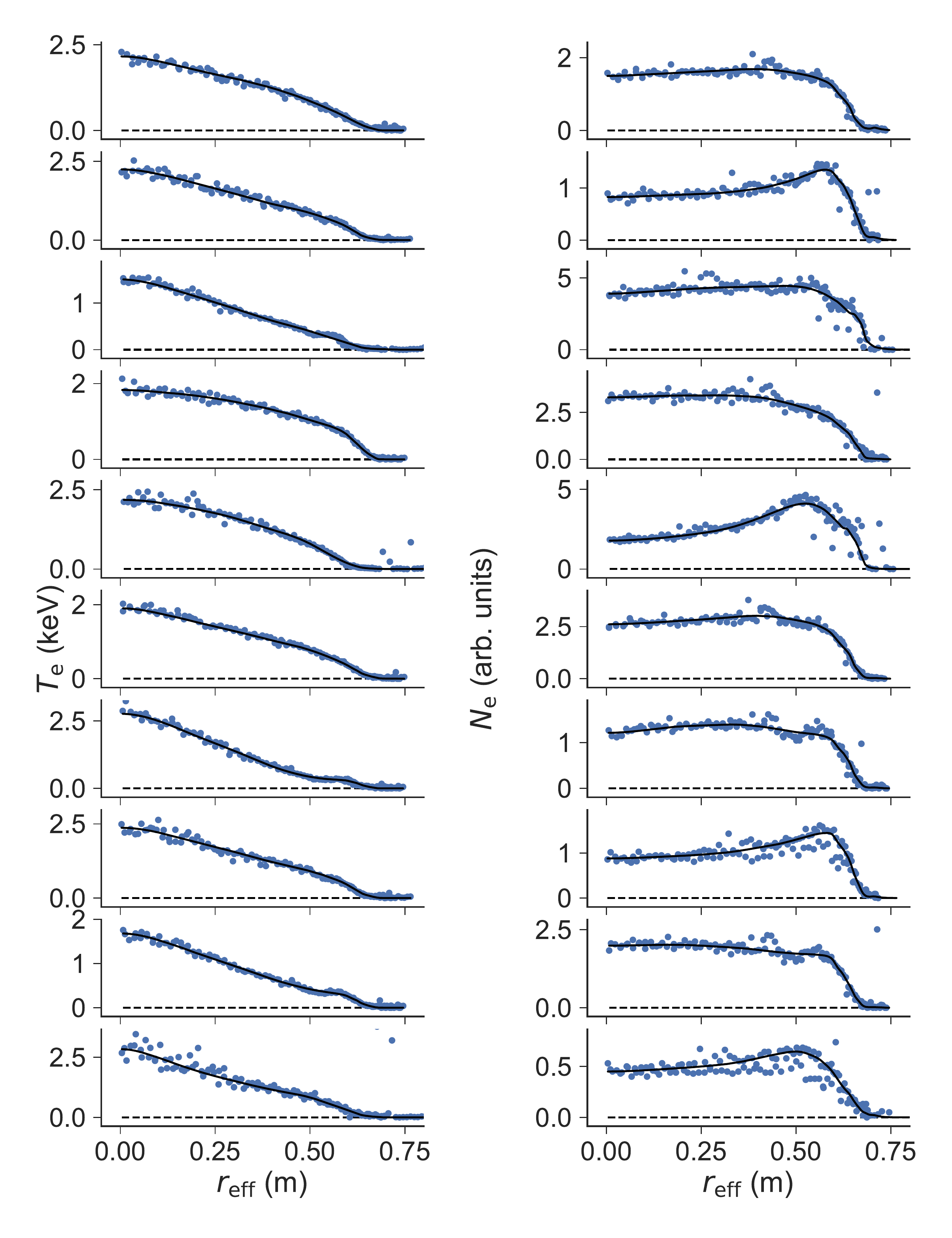}
\end{center}
\caption{
\label{fig:many_fit}
Ten samples randomly chosen from the validation data set.
The figures on the left show $T_\mathrm{e}$ and those on the right show $N_\mathrm{e}$
data (markers). The fit results with our model are shown by the solid curves.
}
\end{figure*}

Figure \ref{fig:many_fit} shows ten random samples from the testing data sets.
It can be seen that the peak values of $T_\mathrm{e}$ distributes
in the range 2--7 keV, whereas $N_\mathrm{e}$ ranges by more than one order.
The fit results by our model are shown by the solid curves in this figure.
The curves nicely fit the data points for all the samples,
even though certain samples include significant outliers,
particularly in the edge region of the plasma ($r_\mathrm{eff} >$ 0.7 m).

\section{Summary and Discussion}

We have developed a method to learn a robust regression algorithm
from the data.
In our method, a generative model is trained that minimizes the KL divergence
from the true data distribution.
We demonstrated that our fitting model outperforms
conventional robust analysis methods
in terms of stability and robustness.

The model developed and trained in this work has already been integrated in the
LHD automatic analysis system.
When new Thomson scattering data arrives, our program runs automatically
and provides the fitting result based on the trained model,
where the global parameters ($\Theta$) are kept fixed
and only the local parameters ($\theta$) are fitted to the data.

\begin{figure*}
\centering
\includegraphics[scale=1.0]{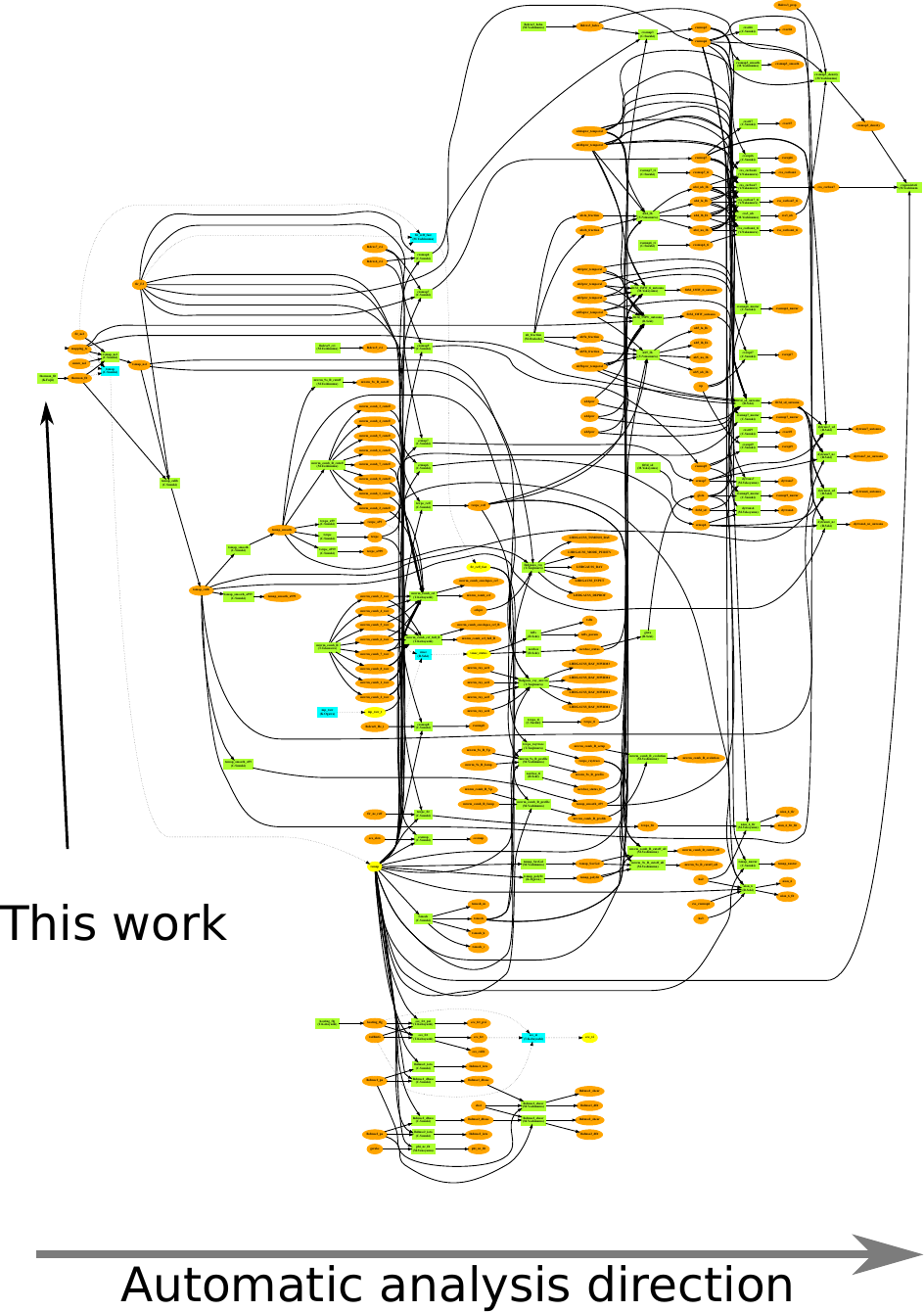}
\caption{
\label{fig:dataflow}
Dependency graph of the LHD automatic analysis system.
Green boxes indicate analysis programs and orange ellipsoids indicate
measurement or analyzed results.
The data flow is represented by black arrows.
The left-most box is the fitting program developed in this work.
Our result is used by more than 80 other programs (green boxes) in the downstream analysis,
and affects more than 100 downstreamed analysis results (orange ellipsoids).
}
\end{figure*}

The results are used by many other automatic analysis programs in LHD.
Figure \ref{fig:dataflow} shows a dependency diagram of the LHD automatic
analysis system.
The green boxes indicate various analysis programs,
and the orange ellipsoids indicate
data generated by measurement or analysis programs.
Our fitting program is placed in the left-most position,
and more than 80 other analysis programs use our results.

We note that our method has a clear limitation,
where we assume that future data distribution can be estimated from
data that have been already obtained.
Therefore, this model will not predict reasonable result
if significant changes in the data distribution occur,
such as
if we change the instrumental hardware (changes in the noise distribution)
or if we become able to produce plasma with extreme parameters
that have not yet been observed.
In order to avoid the effect of the discrepancy between the generative model
and the data distribution,
it may be necessary to train the model continously.

\section*{Acknowledgement}
This work is partly supported by Yazaki Memorial Foundation for Science and Technology, and
the grant of Joint Research by the National
Institutes of Natural Sciences (NINS).

\bibliography{refs}

\appendix
\section{Approximations for Equation \ref{eq:training}
\label{app:approximation}}

There are several difficulties in directly optimizing Eq. \ref{eq:training}.
\begin{enumerate}
  \item \label{item:integration}
  The integration
  $\int
  p(\mathbf{y}^{(k)} |
    \mathbf{f}(\theta | \Theta^\mathrm{fun}), \Theta^\mathrm{lik})\;
  p(\theta | \Theta^\mathrm{pri}) \;
  d \theta$
  is not numerically tractable.
  \item \label{item:large_data}
  The optimization should be done with a large dataset
  $N^\mathrm{train} \approx 10^5$.
\end{enumerate}
In this appendix, we describe the approximations we have adopted in this work.

\subsection{Variational approximation \label{subsubsec:vi}}
In order to approximate the integration,
we adopted the variational approximation \cite{Jordan1999}.
Under this approximation, the posterior distribution of $\theta$ is approximated
by a tractable distribution $q(\theta)$,
from which we can easily sample.
Here, we adopted an independent normal distribution,
\begin{equation}
  p(\theta | \mathbf{y}^{(k)},
  \Theta^\mathrm{fun}, \Theta^\mathrm{lik}, \Theta^\mathrm{pri})
  \approx q(\theta) = \mathcal{N}(\theta | \mu^{(k)}, \sigma^{(k)}),
\end{equation}
where $\mu^{(k)}$ and $\sigma^{(k)}$ are the variational parameter.
This integration bounded from above,
\begin{widetext}
\begin{align}
  & \log\int
  p(\mathbf{y}^{(k)} |
    \mathbf{f}(\theta | \Theta^\mathrm{fun}), \Theta^\mathrm{lik})\;
  p(\theta | \Theta^\mathrm{pri}) \;
  d \theta \\
  & \ge
  \int q(\theta) \;
  \log p(\mathbf{y}^{(k)} |
    \mathbf{f}(\theta | \Theta^\mathrm{fun}), \Theta^\mathrm{lik}) \;
  d \theta
  - \mathrm{KL}\bigl[q(\theta) || p(\theta | \Theta^\mathrm{pri})\bigr]\\
  & \approx
  \log p(\mathbf{y}^{(k)} |
    \mathbf{f}(\theta^q | \Theta^\mathrm{fun}), \Theta^\mathrm{lik})
  - \log(q(\theta^q)) + \log p(\theta^q | \Theta^\mathrm{pri}),
\end{align}
\end{widetext}
where, in the third line, we adopted the Monte Carlo integration with
a sample from $q(\theta)$, indicated by $\theta^q$.

\subsection{Neural network approximation}
Even with the variational approximation,
it is still difficult to optimize the model parameters
($\Theta^\mathrm{fun}, \Theta^\mathrm{lik}$, and $ \Theta^\mathrm{pri}$)
as well as the variational parameters ($\mu^{(k)}$ and $\sigma^{(k)}$),
because of the large number of training data and variational parameters.

We therefore adopted so-called \textit{amortized variational inference}
\cite{Rezende2015},
where the inference process of $\mu^{(k)}$ and $\sigma^{(k)}$ from
$\mathbf{y}^{(k)}$ is approximated by another neural network,
\begin{equation}
  \mu^{(k)}, \sigma^{(k)} \approx N.N.(\mathbf{y}^{(k)} | \Theta^\mathrm{inf})
\end{equation}
where $\Theta^\mathrm{inf}$ is the weight of this network.

With this approximation, we can remove the local parameter $\theta$ from
the optimization target,
and we only need to optimize the grobal parameters
$\Theta^\mathrm{fun}, \Theta^\mathrm{lik}, \Theta^\mathrm{pri}$ and
$\Theta^\mathrm{inf}$.
This enables us to use a stochastic optimization,
where we only use a small part of the data for one iteration to optimize
these grobal parameters,
and in the next iteration, another part of the data is used.
\section{Closeness among the reduced distributions \label{app:closeness}}
In this appendix, we show that
$\mathrm{KL}\bigl[p(\theta) \bigm\vert\bigm\vert p(\theta | \mathcal{M}) \bigr]$
and
$\int\mathrm{KL}\bigl[p(\mathbf{y} | \mathbf{f}) \bigm\vert\bigm\vert
                      p(\mathbf{y} | \mathbf{f}, \mathcal{M}) \bigr]
 d\mathbf{f}$
are the lower bounds of
$\mathrm{KL}\bigl[
  p(\mathbf{y}) \bigm\vert\bigm\vert p(\mathbf{y} | \mathcal{M}) \bigr]$
(Eq. \ref{eq:data_kl}).

Let us consider two different distributions on $\mathbf{y}$,
$p(\mathbf{y})$ and $p(\mathbf{y} | \mathcal{M})$.
If another random variable $\theta$ has a conditional distribution with $\mathbf{y}$,
i.e., $p(\theta|\mathbf{y}, \mathcal{M})$,
the KL-divergence between $p(\mathbf{y})$ and $p(\mathbf{y}|\mathcal{M})$ and
that between their joint distributions
$p(\theta,\mathbf{y}) = p(\theta|\mathbf{y}, \mathcal{M})p(\mathbf{y})$ and
$p(\theta,\mathbf{y} | \mathcal{M}) = p(\theta|\mathbf{y}, \mathcal{M})p(\mathbf{y} | \mathcal{M})$
have the following relation:
\begin{widetext}
\begin{equation}
  \label{eq:kl_joint}
  \mathrm{KL}\bigl[p(\mathbf{y}) \vert\vert p(\mathbf{y}|\mathcal{M}) \bigr]
  = \mathrm{KL}\bigl[p(\theta, \mathbf{y}) \vert\vert p(\theta, \mathbf{y} | \mathcal{M}) \bigr]
  = \mathrm{KL}\bigl[p(\mathbf{y}|\theta)p(\theta)
                     \vert\vert p(\mathbf{y}|\theta, \mathcal{M})p(\theta| \mathcal{M}) \bigr],
\end{equation}
where
$p(\mathbf{y}|\theta) = \frac{p(\theta, \mathbf{y})}{p(\theta)}$ and
$p(\mathbf{y}|\theta, \mathcal{M}) = \frac{p(\theta, \mathbf{y} | \mathcal{M})}{p(\theta | \mathcal{M})}$.
Eq. \ref{eq:kl_joint} can be reduced as follows:
\begin{equation}
\mathrm{KL}\bigl[p(\mathbf{y}|\theta)p(\theta) \vert\vert
                 p(\mathbf{y}|\theta, \mathcal{M})p(\theta | \mathcal{M}) \bigr]
=
\int p(\theta) \mathrm{KL}\bigl[p(\mathbf{y}|\theta) \vert\vert
                                p(\mathbf{y}|\theta, \mathcal{M})\bigr] d\theta
+
\mathrm{KL}\bigl[p(\theta) || p(\theta| \mathcal{M})].
\end{equation}
\end{widetext}

\end{document}